\documentclass[%
aip,
amsmath,amssymb,
reprint,%
floatfix
]{revtex4-2}

\usepackage{graphicx,array,booktabs,rotating}% Include figure files
\usepackage{dcolumn}% Align table columns on decimal point
\usepackage{bm}% bold math
\usepackage{mathptmx}
\usepackage{float,color,xcolor}
\usepackage{hyperref}% add hypertext capabilities
\usepackage{etoolbox} % for \appto
\usepackage{tabularx,tabulary}
\usepackage[capitalize]{cleveref}
\usepackage{times}
\usepackage{textcomp,fontenc}
\hypersetup{
	colorlinks,
	linkcolor={blue!100!black},
	citecolor={blue!100!black},
	urlcolor={blue!100!black}
}
\newcommand{\sk}[1] {\textcolor{black}{#1}}
\newcommand{\sks}[1] {\textcolor{black}{#1}}
\newcommand{\etal}{\textit{et al.}}

\usepackage[pagewise,mathlines]{lineno} % Enable numbering of text and display math
\begin{document}
	
	\preprint{AIP/123-QED}
	
	%\title[Sample title]{Sample Title:\\with Forced Linebreak\footnote{Error!}}% Force line breaks with \\
	%\thanks{Footnote to title of article.}
	\title[Phys. Fluids (2020): Revised Manuscript Ver. 2.0]{On the fluidic behavior of an over-expanded planar plug nozzle under lateral confinement}
	%\author{A. Author}
	\author{M. Chaudhary}%
	\affiliation{Department of Aerospace Engineering, Indian Institute of Technology, Kanpur-208016, India}%
	
	\author{T. V. Krishna}%
	\affiliation{Department of Aerospace Engineering, Indian Institute of Technology, Kanpur-208016, India}%
	
	\author{Sowmya R. Nanda}%
	\affiliation{Department of Aerospace Engineering, Indian Institute of Technology, Kanpur-208016, India}%
	
	\author{S. K. Karthick}
	%\homepage{http://www.Second.institution.edu/~Charlie.Author.}
	\affiliation{Faculty of Aerospace Engineering, Technion-Israel Institute of Technology, Haifa-3200003, Israel}%
	
	\author{A. Khan}%
	\affiliation{Department of Aerospace Engineering, Indian Institute of Technology, Kanpur-208016, India}%
	
	\author{A. De}%
	\affiliation{Department of Aerospace Engineering, Indian Institute of Technology, Kanpur-208016, India}%
	
	\author{Ibrahim M. Sugarno}
	\email{Ibrahim@iitk.ac.in}
	\affiliation{Department of Aerospace Engineering, Indian Institute of Technology, Kanpur-208016, India}%
	\date{\today}% It is always \today, today,
	
	\begin{abstract}
		The present work aims to study the fluidic behavior on lateral confinement by placing side-walls on the planar plug nozzle through experiments. The study involves two cases of nozzle pressure ratio (NPR=3, 6), which correspond to over-expanded nozzle operating conditions. Steady-state pressure measurements, together with schlieren and surface oil flow visualization, reveal the presence of over-expansion shock and subsequent interaction and modification of the flow field on the plug surface. The flow remains attached to the plug surface for NPR=3; whereas, for NPR=6, a separated flow field with a recirculation bubble is observed. Spectral analysis of the unsteady pressure signals illustrates a clear difference between the attached and the separated flow. Besides, other flow features with a distinct temporal mode associated with and without lateral confinement are observed. The absence of lateral confinement reduces the intensity of low-frequency unsteadiness; however, on the contrary, the interaction region is relatively reduced under lateral confinement.
	\end{abstract}
	
	%Use showkeys class option if keyword
	\maketitle
	\section{Introduction}\label{introduction}
	A conventional convergent-divergent (C-D) nozzle, though successfully used in propulsion systems of rockets and launch vehicles, undergoes performance losses when operating at highly over-expanded conditions. These fixed geometry nozzles are generally designed for a particular Mach number ($M_d$), hence giving optimal efficiency at a particular nozzle pressure ratio ($p_0/p$). \sk{However, the pressure changes continuously with the altitude as the vehicle ascends through the atmosphere, thereby making the nozzle to operate at both over-expanded and under-expanded regime.} To overcome the losses encountered in the over-expanded regime and to enhance the efficiency of the propulsion system, the nozzle should be altitude adaptable. Variable throat C-D nozzle is one option; however, they are costly and difficult to manufacture. Besides, it requires constant control for its operation at varying altitudes. The factors mentioned above eventually led the researchers to explore the feasibility of using plug nozzles for rocket/missile applications.
	
	Plug nozzles were used in earlier turbojets and for airplane applications before the Second World War \cite{AUKERMAN1991,Sutton2006}. The concept of plug nozzles and its \sk{performance characteristics} for rocket propulsion was first suggested by Griffith \cite{Griffith1954}. Plug nozzle provides altitude compensation with improved nozzle efficiency both at low altitude and high altitude conditions as compared to conventional bell nozzle \cite{Hagemann1998,Ruf_1997}. Different groups studied \sk{the overall flow features of plug nozzle} across the globe in the past and provided in-depth insight into the performance of plug nozzle \cite{Onofri2002,Chutkey2012}. Full-length plug nozzle usually adds significant weight to the vehicle, so truncated plug nozzle is usually incorporated to gain net performance as the only first quarter of plug nozzle produces significant thrust. This loss of thrust is compensated by the nozzle base pressure that acts on the truncated base area \cite{Ruf_1997,Tomita1996}.
	
	\sk{One of the major problems of conventional C-D nozzle is the over-expanded nozzle operation at a pressure ratio below its design point\cite{Romine1998,Zebiri2020} (low altitude). Such an operation forms a shock wave inside the nozzle, and the flow downstream gets separated from the nozzle wall\cite{Hamed1998}. The flow separation makes the flow severely unsteady, leading to the origin of fluctuating side loads and structural vibration\cite{Ostlund2005}. In internal flows like the jet in confinement or shrouded jet \cite{Gupta2019,Karthick_2016}, these features are observed to be beneficial in reducing jet noise, minimizing thermal signature of the jet, and enhancing fluid mixing. Some of the researchers\cite{Kumar2018,ArunKumar2019a,ArunKumar2019b} exploit the off-design behavior using active or passive methods to further control the jet. However, the flow-induced vibrations can reduce the overall performance of the nozzle and may even result in engine failure in extreme conditions \cite{Frey1999,Verma_2008,Johnson2010,Olson_2013}. Researchers \cite{Lawrence1967,Reijasse_1999,Bourgoing_2005,Papamoschou2008,Verma_2014, Verma2014} had performed a large number of experimental studies to explore the cause of flow unsteadiness, and some of the investigations are still in process.}
	
	Planar plug nozzles are also subjected to over-expansion conditions when operated at nozzle pressure ratio (NPR$=p_0/p_a$) lower than the value corresponding to the designed condition ($M_d$). In these cases, shock waves occur on the plug surface leading to flow unsteadiness. Unlike the annular plug nozzles, where the flow expands axisymmetrically, the planar plug nozzles have flow expansion in the streamwise direction. The process of expansion can also occur in the spanwise direction in the absence of side-walls. Such lateral expansion of the flow leads to thrust losses. Miyamoto \etal \cite{Miyamoto_2006} studied the effects of side-wall and side-wall edge angle and reported thrust loss in the absence of the side-wall. Also, the side-wall edge angle neither affects the flow field nor the thrust performance, primarily.  Other researchers \cite{Ito_2004,Mori} report similar effects on the thrust performance due to side-wall in a linear plug nozzle. Verma and Viji \cite{Verma_2011} assessed the effect of the side-wall and free stream flow on the base pressure measurement for 40\% truncated contoured and conical plug nozzle. The complex interaction of over-expansion and recompression shocks on the plug surface causes flow unsteadiness. Researchers perceive such events as a potential source for structural damage. They also attempted to identify the source of unsteadiness in the exit flow and the parameters governing the base pressure unsteadiness through quantification of shock oscillation frequency. Chutkey \etal \cite{Chutkey_2017} also reported unsteadiness associated with the flow field of the linear plug nozzle and the effect of clustering. \sks{However, a proper understanding of the unsteady flow events occurring on the plug surface due to over-expansion shock-induced flow separation, and the effects of adding side-wall to the plug nozzle on the gross flow field is not clear till date.}
	
	Therefore, in the purview of the literature, it raises few questions \sk{as: What happens to the flow field when the strength of the over-expansion shock increases by increasing NPR? Will} the unsteadiness and temporal characteristics remain the same or \sks{differ}?  How does the side-wall change/influence the flow field characteristics on the plug surface? These queries are addressed in the present work with the help of both qualitative and quantitative measurements.
	
	Experiments are carried out on a half planar plug nozzle configuration at two different NPR's, for two different types of flows, an attached and a separated flow, corresponding to the over-expanded case of nozzle operation. The effect of lateral confinement is studied by attaching side plates until the full length of the plug nozzle. The influence of side-wall and the effect of NPR's on the nozzle flow field are visualized using schlieren and oil flow visualization techniques to extract qualitative features of the flow. Detailed flow dynamics are studied by measuring time-averaged static pressure on the plug surface at both the center-line and close to the side end of the nozzle. Furthermore, unsteady pressure measurements are carried out to identify the dominant temporal characteristics. 
	
	The paper is organized as follows. In Section \ref{exp_setup} and \ref{num_method}, the experimental and numerical methodology is discussed. In Section \ref{res_disc}, qualitative results from the schlieren and the oil flow are presented first. Steady and unsteady pressure measurements are used later to make quantitative remarks. Computational results are used as supplements in each of the sections to understand the flow physics. In Section \ref{conclusions}, vital conclusions from the present study are listed.
	
	 \section{Experimental Methodology}\label{exp_setup}
	
	\begin{figure*}
		\centering
		\includegraphics[width=0.7\textwidth]{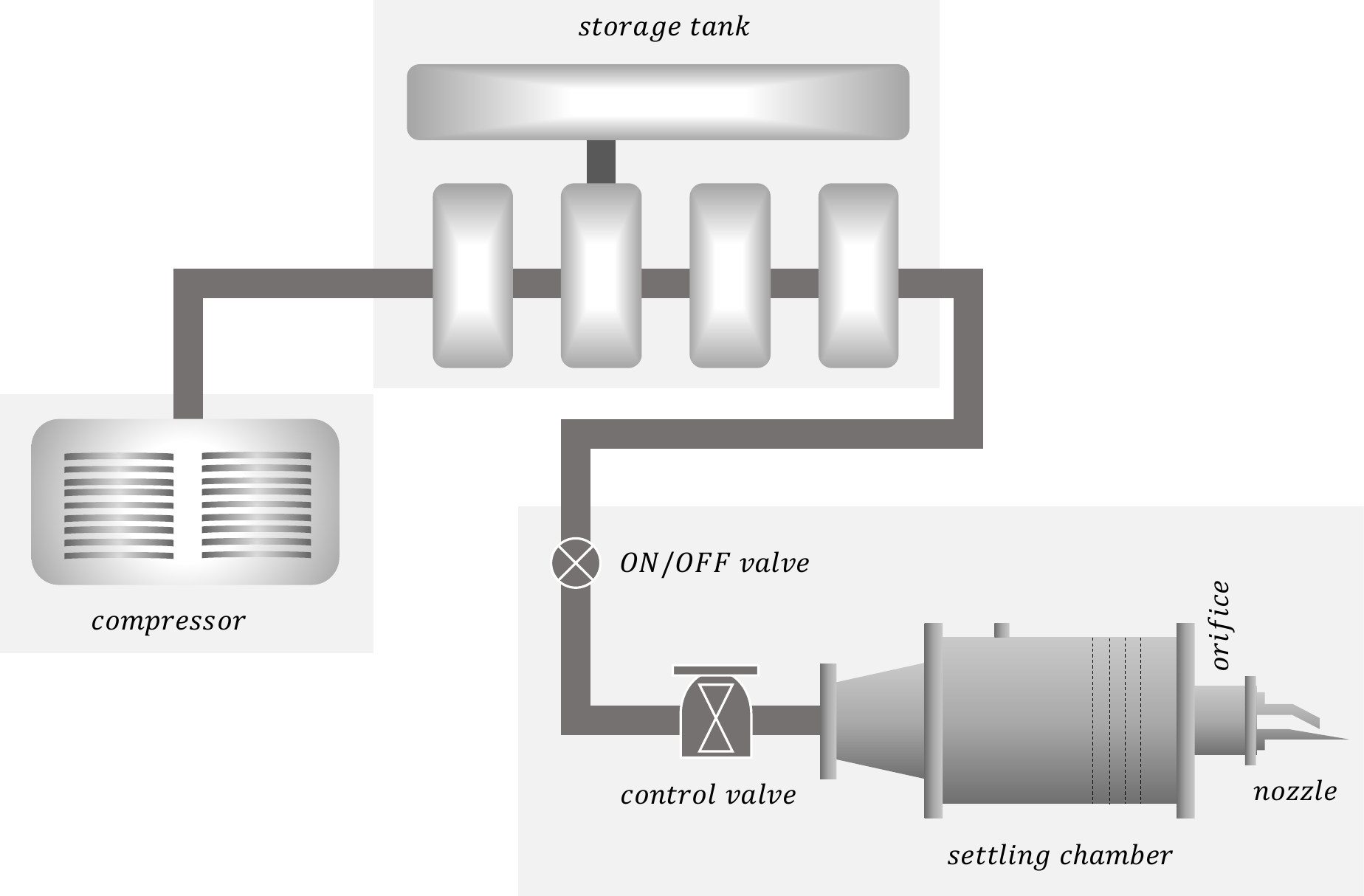}
		\caption{A schematic showing the layout of the experimental facility where the planar plug nozzle studies are carried out.}\label{fig1}
	\end{figure*}
	
	\subsection{Free jet facility}
	Experiments are conducted in the free jet facility at High-Speed Aerodynamics Laboratory, Indian Institute of Technology Kanpur, India \cite{Khan_2019}. A schematic of the setup is shown in Figure \ref{fig1}. The facility consists of a multi-stage reciprocating compressor to charge the decontaminated compressed air in the multiple storage tanks. The maximum storage pressure in these tanks is 3.5 MPa. Later, the compressed air from the storage tanks is supplied to the settling chamber through the control valves. Various mesh-screens and honeycombs are positioned inside the settling chamber for flow straightening and conditioning. The designed plug nozzle is connected to the orifice at the end of the stagnation chamber, as shown in Figure \ref{fig1}.
	
	\begin{figure*}
		\centering
		\includegraphics[width=1\textwidth]{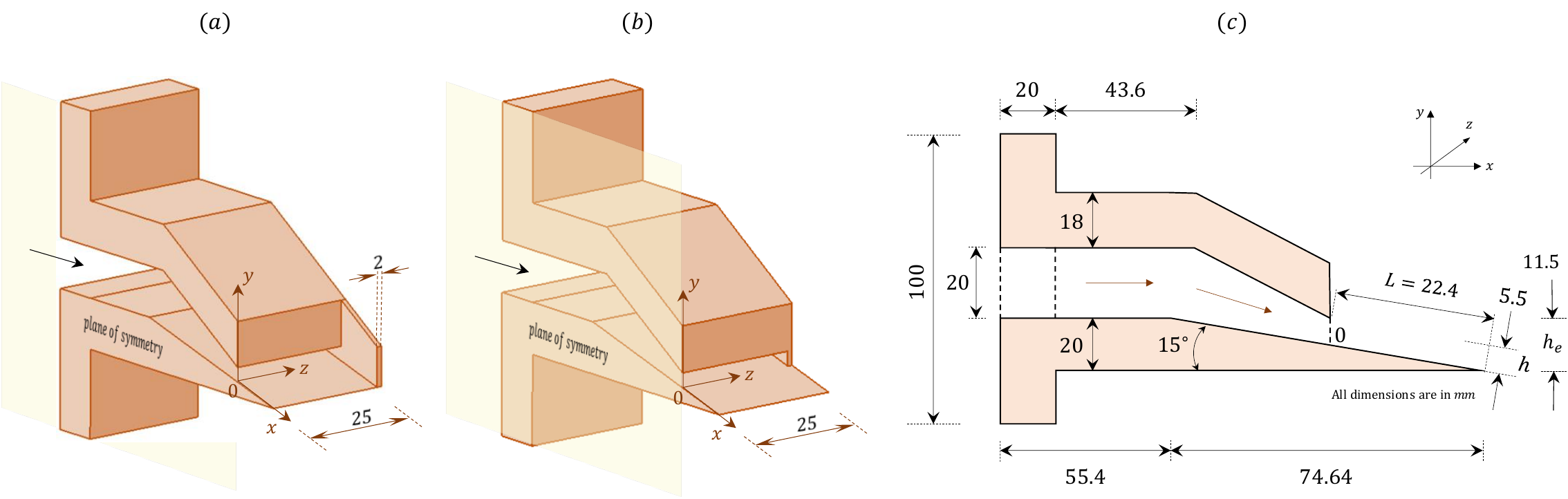}
		\caption{Isometric view of the planar plug nozzle (a) with and (b) without side-walls shown about the plane of symmetry ($xy$-plane). (c) Basic geometrical details of the planar plug nozzle. Co-ordinate system description: $x$-streamwise, $y$-transverse, and $z$-spanwise or lateral direction. $o$-origin, $h$-height, $w$-width, and $L$-length. All dimensions are in mm.}\label{fig2}
	\end{figure*}
	
	\begin{table}
		\caption{List of jet parameters\footnote{$p_0/p$-nozzle pressure ratio (NPR), $T$-temperature, $u$-streamwise velocity, $\nu$-kinematic viscosity, $h$-height, $M$-Mach number, $Re$-Reynolds number. Subscripts: $j$-fully-expanded jet condition, $0$-total condition.} realized during the experiments and computations at different nozzle pressure ratio (NPR=$p_0/p$) for a constant total temperature of $T_0=300$ K, \sk{throat streamwise velocity of $u^*=316.94$ m/s}, \sks{and apparent jet exit height of $h_e=11.5$ mm.}}
		\label{table:flow_cond}
		\begin{ruledtabular}
			\begin{tabular}{@{}c c c c c c@{}}
				$\left[p_0/p\right]$ &  $u_j$ (m/s) & $\nu_j \times 10^5$(m$^2$/s) & $h_j$ (mm) & $M_j$ & $Re_j(h_j)$ \\
				\midrule
				3 &    402.95 &    0.89 &    6.01 &    1.36 &    2.69\\
				6 &    491.41 &    0.62 &    8.09 &    1.83 &    6.42\\
			\end{tabular}
		\end{ruledtabular}
	\end{table}
	
	\subsection{Experimental Model}
	A schematic of the plug nozzle with and without side-walls are shown in Figure \ref{fig2}. The model has a throat height  $ h = 5.5 $ mm, slant length $ L = 22.4 $ mm, width $ w = 50 $ mm and plug half-angle of $ 15^\circ $. Two different lengths of side plates are used for studying the effect of the side-wall, with one side plate running up to the throat and another side plate till the full-length plug. In total, seven wall-static pressure sensors were placed along the streamwise direction with an axial offset of 3 mm between $ [x/L]=0 $ and $ [x/L]=0.8 $. A linear array of the sensors, as mentioned above, is available along the center-line $ [z/L]=0 $ and also close to the side-wall $ [z/L]=1 $ of the planar plug nozzle to compare the effect of the side-wall on the flow field. Typical flow conditions realized during the nozzle run are summarized in Table \ref{table:flow_cond}.
	
	\subsection{Flow Visualization (Schlieren and oil flow)}
	In the present work, flow visualization is carried out through a `Z-type' schlieren system \cite{Settles2001} for the case without the side-wall as the other case offer challenges in terms of optical access. Schlieren system employs a 3W white LED light source with a power supply (Model: HO-HBL-3W, Holmarc), and a high-speed camera (Model: CH14-1.0-32C, Chronos) for capturing the images along with the appropriate optics. Schlieren mirrors are 200 mm in diameter with a focal length of 1.5 m. The images are captured at a sampling frequency of $ f_s=1000 $ Hz with \sks{a light/frame} exposure time of  $t_e =100$ $\mu s$. The frame resolution is 600$\times$300 pixels at a spatial resolution of 0.1618 mm/pixel. Spatial fields are scaled using the slant length of the plug surface $(L)$. Surface flow visualization is also performed with the help of an oil flow visualization technique to compare the flow features of the plug nozzle with and without a side-wall. A mixture of titanium dioxide, oleic acid, and vacuum oil is used for oil flow visualization. A black acrylic sheet of 0.1 mm thickness is pasted on the nozzle plug surface, and the mixture is sprayed on the plug surface to get an insight on the flow feature during the test time. A Nikon\textsuperscript{\tiny\textregistered} D70 DSLR camera with 16 mega-pixel resolution is used to capture the flow field with a spatial resolution of 0.082 mm/pixel.
	
	\begin{figure*}
		\centering
		\includegraphics[width=0.8\textwidth]{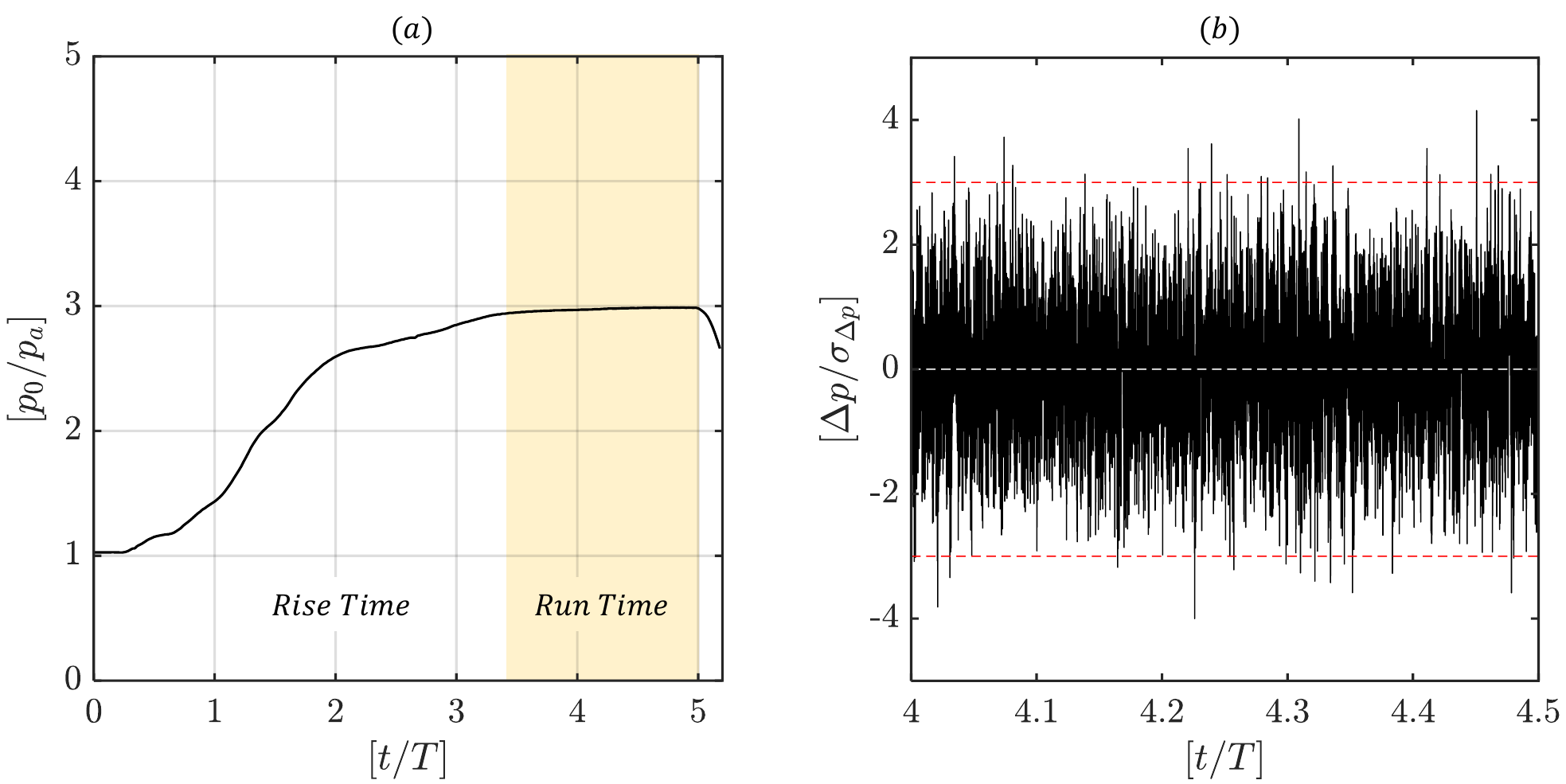}
		\caption{(a) A typical pressure signal obtained from a pressure sensor mounted on the stagnation chamber shows the variation of the non-dimensional pressure $ (p_0/p_a) $ with respect to the non-dimensionalized flow time ($t/T$, where $T=1$ s) for NPR=3. Steady run time achieved during the test is shown in pale-yellow color. (b)  Variations observed from a typical non-dimensionalized differential unsteady pressure signal ($\Delta p$) from the PCB\textsuperscript{\tiny\textregistered} sensor mounted on the planar plug nozzle surface along the center-line at the location $[x/L=0.8]$ and $[z/L=0]$ during the steady run time $ (4.0\leq [t/T]\leq 4.5) $. The dotted red-line marks the deviation bounds at $ \pm 3\sigma $ and the dotted white-line marks the zero line.}\label{fig3}
	\end{figure*}
	
	\subsection{Data acquisition system and pressure measurements (steady and unsteady)}
	Steady pressure measurements are acquired on the center and side-line of the plug surface using a 16-channel steady pressure scanner (Pressure Systems, Inc., Model-9016). One channel of the pressure scanner is used to measure the stagnation chamber pressure, and others are used to measure the wall-static pressure on the plug surface. A data acquisition system is used to acquire the data at a sampling rate of 50 Hz over a period of 1 s. Unsteady pressure measurements on the plug surface are carried out using PCB\textsuperscript{\tiny\textregistered} piezotronics pressure sensors (Model-113B24) mounted on the center-line and side-line of the plug surface at selected locations of interest$ ([x/L]=0.8) $. The data are acquired at a sampling rate of 100 kHz for a duration of 1 s. A typical steady and unsteady pressure signal achieved during the experiments is shown in Figure \ref{fig3}, where the experimental run-time is marked.
	
	\subsection{Uncertainty}
	The sources of uncertainty in the measured and the derived quantities are calculated as per the procedures mentioned in the book of Coleman and Steele \cite{Coleman2009}. Uncertainty in the measured quantities for the steady and unsteady pressure sensors is calculated to be $\pm$3\% and $\pm$10\% about the measured values. Spatial uncertainty in imaging the flow features equals the pixel resolution itself.  Thereby, schlieren and oil flow visualization comprises features having a spatial variation of 0.2 mm and 0.1 mm, respectively. The spectral resolution from the unsteady measurements is about 49 Hz.
	
	\section{Numerical Methodology} \label{num_method}
\begin{figure*}
		\centering
		\includegraphics[width=0.9\textwidth]{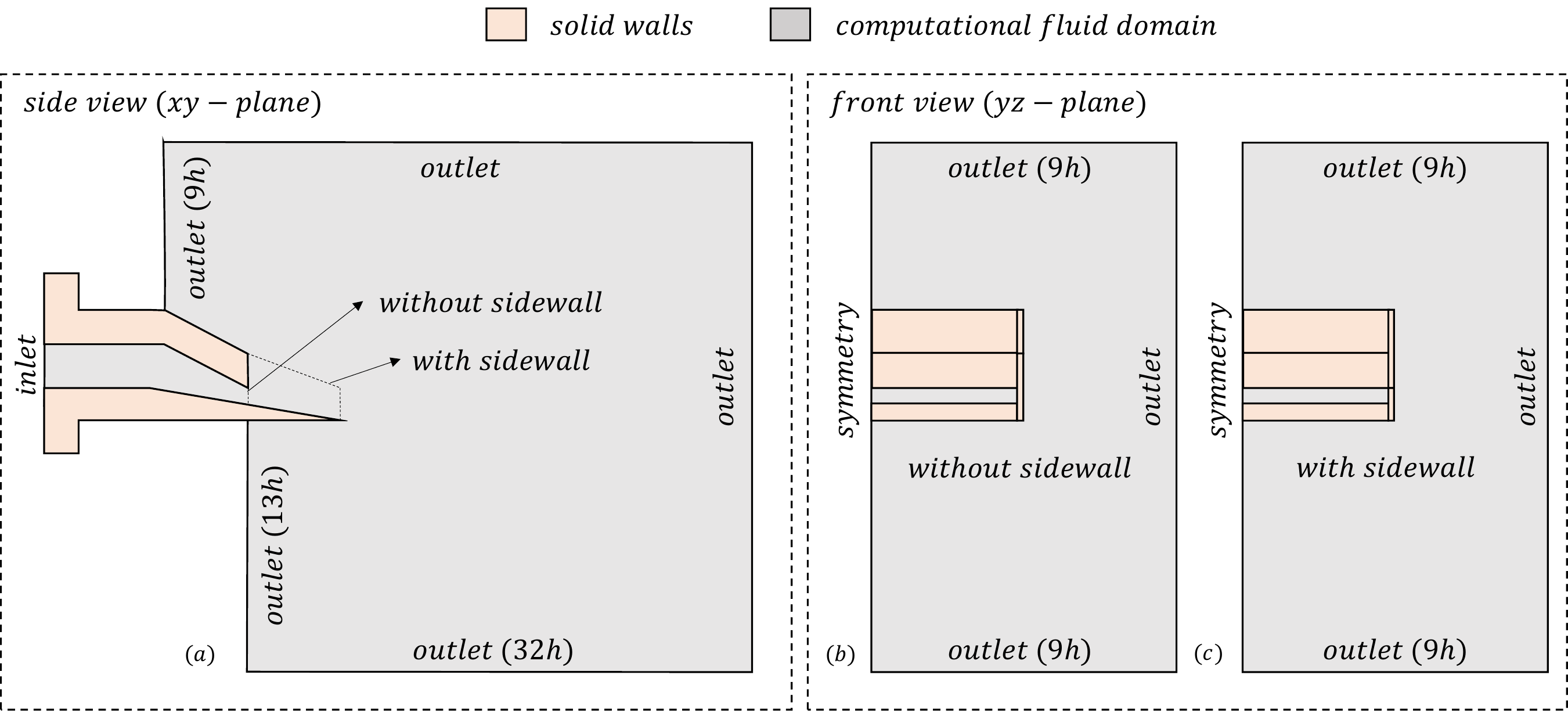}
		\caption{Schematic showing the side view and front view of the computational domain along with the boundary conditions. (a) Domain extents used in the 2D-RANS/3D-RANS along the $ xy $-plane, (b-c) Domain extents for ramp nozzle simulations without and with side-walls along the $ yz $-plane.}\label{fig4}
	\end{figure*}

\begin{figure*}
		\centering
		\includegraphics[width=1\textwidth]{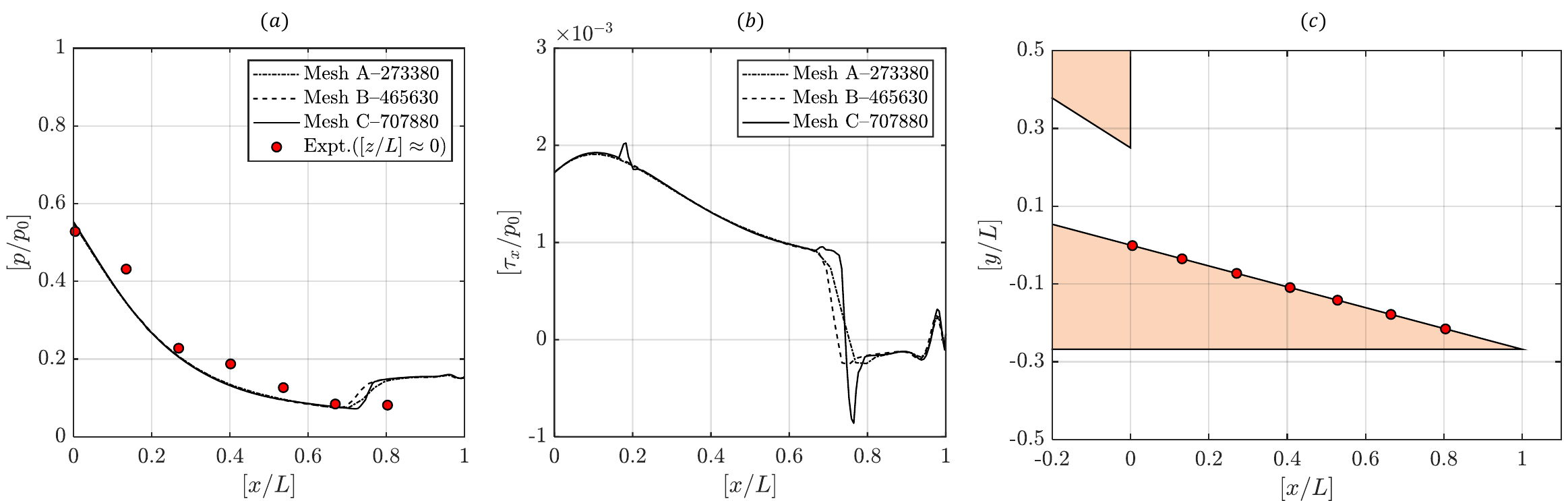}
		\caption{\sk{Results of grid independence studies showing the variations observed along the planar plug nozzle surface between $0 \leq [x/L] \leq 1$ for NPR=6 with sidewalls at $[z/L]=0$: (a) wall-static pressure distribution ($p/p_0$), (b) shear-stress distribution ($\tau_x/p_0$) along the $x$-direction for three different mesh densities, and (c) wall-static pressure measurement locations on the planar plug nozzle surface with and without side-walls at $[z/L]=0$ plane (which is also the same at $[z/L]\approx 1$).}}\label{fig5}
	\end{figure*}

	Numerical simulations are carried out using a commercial computational fluid dynamics (CFD) solver Ansys-Fluent\textsuperscript{\tiny\textregistered} to solve the Reynolds Averaged Navier-Stokes (RANS) equations \cite{Alam_2016,SHIMSHI2009}. The present work involves both two-dimensional and three-dimensional RANS simulations. The CFD domain boundary conditions (wall, symmetry, pressure inlet, and pressure outlet) used in the simulations are also shown in Figure \ref{fig4}. The spatial grids are generated using the Ansys-ICEM\textsuperscript{\tiny\textregistered} module. The computational cells are of structured quadrilateral types, while an equisize skewness values are contained well below 0.2 for 95\% of the cells. \sk{A turbulence model of `$k-\omega$-SST' (shear stress transport) is used with compressibility corrections and standard wall function as it is known to predict the flow separations in the jet flow field \cite{Balabel_2011,SHIMSHI2009}. Near the wall, $y^+$ values are maintained less than 5 \cite{Yu2012} with a cell spacing progression being maintained not larger than 1.2 normal to the wall. Grid independence studies are carried out and the results are shown in Figure \ref{fig5}. A final grid (Mesh-C) of $\sim 0.1$ million for 2D-RANS and $\sim 1$ million for 3D-RANS are found to be sufficient for the present analysis.} In the coupled pressure-based solver, we use air as the ideal gas. While performing spatial discretization, Green-Gauss node-based method is used to resolve the gradients, and a second-order scheme is used to resolve pressure, density, momentum, turbulent kinetic energy, specific dissipation rate, and energy. A pseudo-transient solution steering methodology, along with hybrid initialization, is adopted for rapid convergence of the solution, and convergence of $10^{-5}$ is achieved in the continuity equation in all the performed computations.
	
\begin{figure*}
		\centering
		\includegraphics[width=0.8\textwidth]{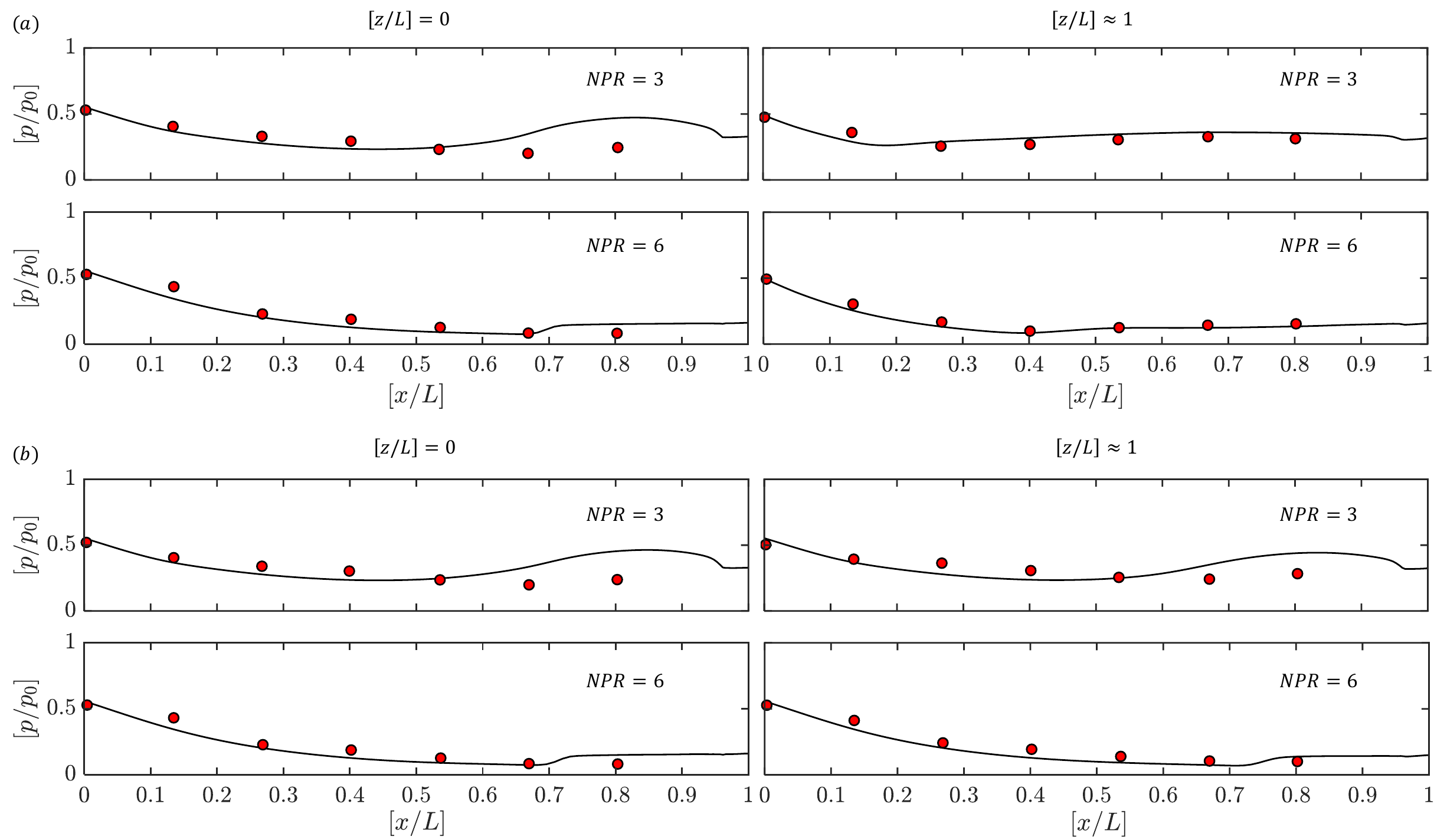}
		\caption{Comparison of experimental and computational normalized wall-static pressure distribution along the planar plug nozzle surface $ (x/L) $ (a) without and (b) with side-walls at two different spanwise locations ($ [z/L]=0 $,and $ [z/L]\approx 1 $). The solid line is from CFD results, and the red-filled solid markers are from the experiments.}\label{fig6}
	\end{figure*}

	Computations are validated against the present experiments through the wall-static pressure measurements on the planar plug surface, as shown in Figure \ref{fig6}. Measurements along the center-line and close to the side-wall are plotted for both the operating conditions with and without lateral confinement. The numerical results are found to be in good agreement with measurements. The deviation seen closer to the end of the planar-plug surface is attributed to the poor spatial resolution in sensor placement and the inherent size of the sensor itself.
	
	\section{Result and Discussions}\label{res_disc}
	The effect of the side-wall on the flow dynamics occurring on the planar plug nozzle surface is studied at two different NPR's (3 and 6), and the jet parameters are tabulated in Table \ref{table:flow_cond}. The two conditions are selected as the flow remains attached to the plug surface at NPR=3 and detaches at NPR=6. Qualitative and quantitative comparisons from the flow visualization through schlieren and oil flow technique together with pressure measurements (both steady and unsteady) on the plug surface shed valuable information. Tandem findings from computational results give a clear understanding of the flow field, and the overall results are explained in the following sections.
	
	\subsection{Schlieren imaging at different NPR's without side-walls}
\begin{figure*}
		\centering
		\includegraphics[width=0.84\textwidth]{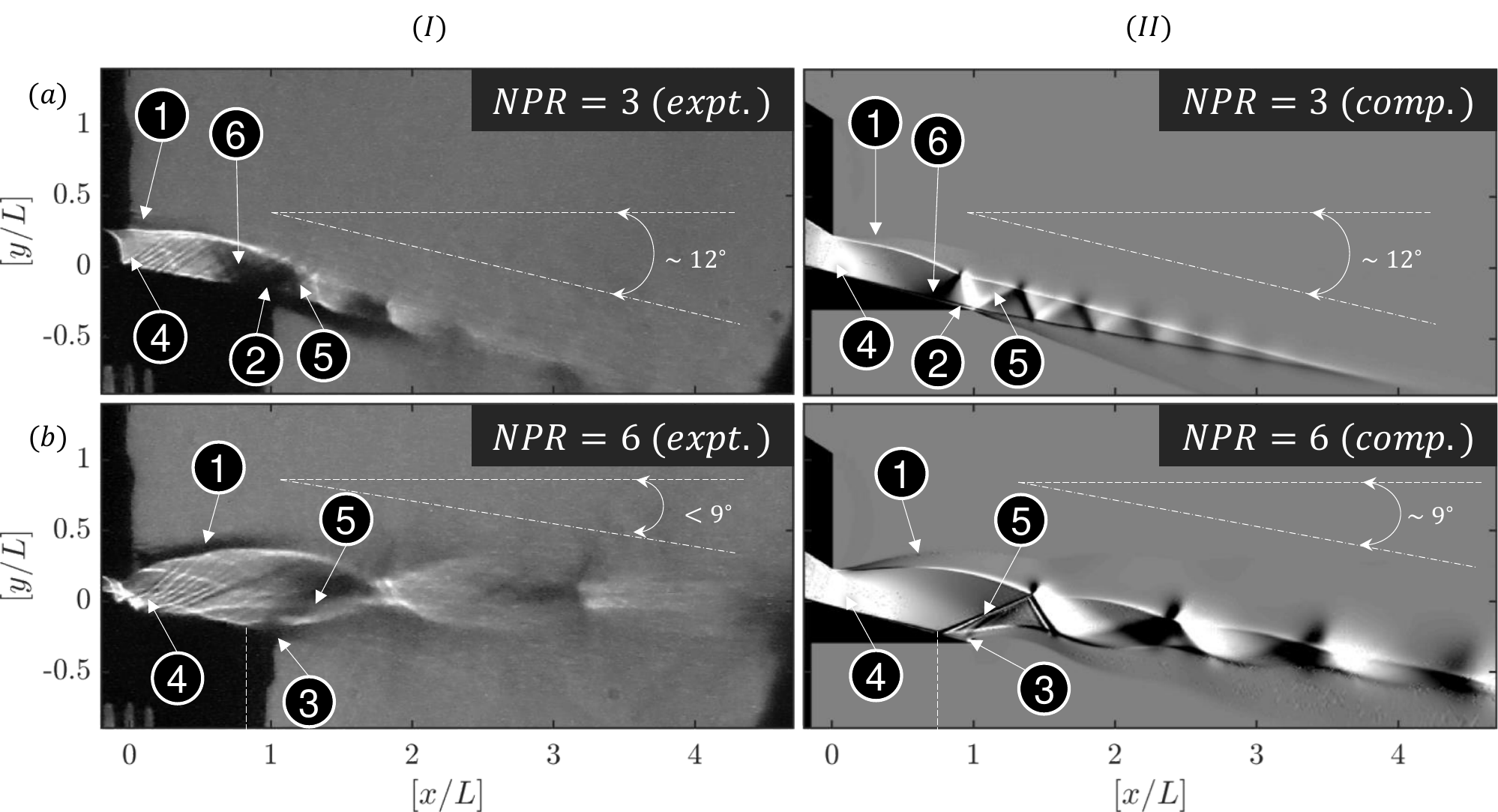}
		\caption{\sks{Typical time-averaged schlieren images obtained from (I) experiments and (II) computations ($||\partial{\rho}/\partial{x}||$, normalized density gradient along the streamwise direction) at two different NPR's (a. NPR=3, and b. NPR=6) for the planar plug nozzle flows without side-walls. Flow features: 1. Free shear layer, 2. Attached flow, 3. Separated flow, 4. Expansion waves, 5. Over-expansion shock, and 6. Reflected compression waves. The point of separation is marked using vertical dotted white line for NPR=6. The difference in jet inclination angle between the experiments and CFD at NPR=6 is due to the imaging limitations of the camera in capturing high-frequency oscillations and low-exposure of light, three-dimensional line of sight light integration, and mainly due to downstream jet-column flapping.}}\label{fig7}
	\end{figure*}

	Figure \ref{fig7} shows the wave structure prevailing in the plug nozzle jet obtained using the schlieren technique. Images on the left side show instantaneous flow field obtained from experiments, whereas, images on the right side show time-averaged flow field generated numerically by calculating the gradients of the density field along the streamwise direction. The comparison is made only for the case without the side-walls presence since it is not possible to obtain experimental schlieren images with the presence of side-wall. A striking similarity between the experimental and the numerical results, especially near the plug surface, is achieved. The pressure distribution on the planar plug nozzle depends on the evolution of the fluid on the plug surface, which is primarily governed by the NPR. The fluid coming from the settling chamber expands internally to sonic state as it reaches the cowl exit. Since the cowl is designed to have a minimum cross-sectional area at the exit, the fluid is choked right at the cowl lip. As the fluid leaves the cowl, it undergoes supersonic expansion on the \sk{plug} surface. This expansion occurs externally since the jet plume on the ramp surface is exposed. The upper jet boundary is free to adjust to the outer atmosphere. Therefore, external expansion nozzles are capable of adapting to the changing ambient conditions resulting in better off-design performance as compared to conventional nozzles.
	
	In Figure \ref{fig7}a, flow field for NPR=3 is shown where fluid expands near the cowl through an expansion fan that is centered on the cow lip. \sks{In Table \ref{table:flow_cond}, the fully expanded jet height ($h_j$) and the apparent jet exit height ($h_e$) are given, where $h_j<h_e$ for both the NPR's}. From the free/confined jet nomenclature \cite{Karthick_2016,Tam_1982}, the farther the values of $ h_j $ from $h_e $, the higher the degree of over-expansion, and it is vice versa if $h_j$ approaches $h_e$. All the NPR cases are identified to be over-expanded, with the degree of over-expansion being higher for NPR=3 than NPR=6. The nozzle at NPR=3 is highly over-expanded because of which the expansion waves (the white region near the cowl-lip) are confined to a small region of the ramp surface near the cowl exit. These expansion waves are reflected from the ramp towards the outer shear layer. These expansion waves then are reflected from the shear layer as weak compression waves and result in the formation of an over-expansion shock wave on the ramp, as shown in Figure \ref{fig7}a. 
	
	The formation of over-expansion shock wave causes a sudden rise in the static pressure distribution on the ramp surface, as evident from Figure \ref{fig6}. The shock wave interacts with the upper shear (white) layer and is reflected as expansion waves towards the lower shear layer (black). The back and forth a reflection of shocks and expansion waves results in the formation of a shock-cell structure that is confined within the supersonic core of the jet. The supersonic core is bounded by the upper and the lower shear layer, as can be seen from the schlieren image (Figure \ref{fig7}a).  The jet remains attached to the ramp surface, and no evidence of flow separation is observed. The boundary of the jet is straight, and the jet itself is aligned with the plug surface with the same inclination of $15 ^\circ$, \sk{as shown in Figure \ref{fig2}c}. After leaving the plug surface, the jet trajectory is inclined to $\sim 12 ^\circ$ with the horizontal plane.
	
	As the NPR is increased to 6, as shown in Figure \ref{fig7}b, the jet expands strongly near the cowl lip. The free shear layer opens up to a width more extensive than that of the NPR=3 cases. As the jet width rapidly increases in the free shear layer side, the fully-expanded jet width $(h_j)$ is realized well before the termination of the plug wall surface. Hence, the flow separates shortly before leaving the plug surface. Because of the separation, the jet boundary turns away from the plug surface through a strong oblique shock, as shown in Figure \ref{fig7}b. \sks{The difference in jet inclination angle between the experiments and CFD at NPR=6 in Figure \ref{fig7}b is due to the imaging limitations of the camera, three-dimensional line of sight light integration and the fact that the jet is flapping (unsteady). However, the gross flow features like the separation point on the plug-surface, jet inclination being smaller than the NPR=3 case are still observed in the image.}
	
	Since the nozzle plug surface is linear and is not designed for smooth expansion, a strong oblique shock wave is inevitable, and it stands near the tip of the plug surface. Due to the strong oblique shock wave encounter, the boundary layer separates from the ramp wall, and the lower shear layer is turned away from the wall. The interaction of the oblique shock wave with the boundary layer is often responsible for the generation of unsteady pressure fluctuations on the wall, which is discussed in the upcoming section. A more extended supersonic core with bigger shock cells in comparison to NPR=3 is observed for NPR=6. After separation at $[x/L] \approx 0.8$, the jet trajectory is inclined at even a lesser angle $(\sim 9^\circ)$ with the horizontal plane than that of the previous.
	
	\subsection{Oil flow visualization}
\begin{figure*}
		\centering
		\includegraphics[width=0.85\textwidth]{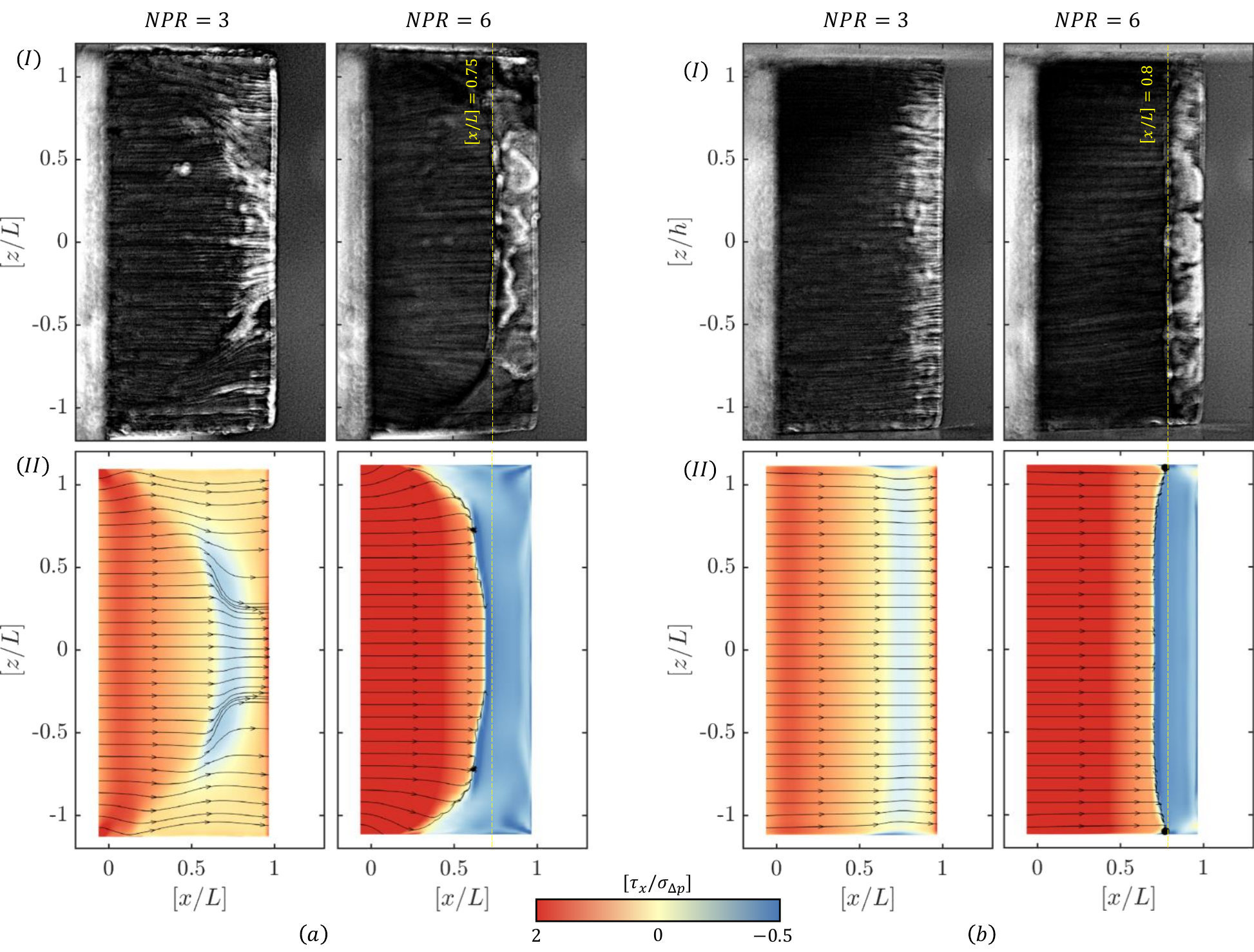}
		\caption{\sks{Comparison of (I) oil flow visualization with the (II) normalized shear-stress distribution ($\tau_x/\sigma_{\Delta p}$) along the $x$-direction from the 3D-RANS simulations for the two cases of NPR (3 and 6) in a planar plug nozzle flows (a) without and (b) with side-walls. A common deviation of $ \sigma_{\Delta p}=0.1 $ kPa is used for the wall-shear normalization to appreciate the comparison between cases.}}\label{fig8}
	\end{figure*}

	The images obtained from the oil flow visualization technique are processed using MATLAB\textsuperscript{\tiny\textregistered} and are compared with the computational results. The values of shear stress ($\tau_x$) along the $x$-direction is calculated using the Equation \ref{eqn} (where $\mu$-dynamic viscosity, $u,v,w$-velocity components in $x,y,z$ directions) mentioned below to compare the numerical results with the experiments. Parallel comparison of streaks from the oil flow visualization with the total streamwise shear stress from the computations is given in Figure \ref{fig8}. 
	
	\begin{equation} \label{eqn}
	\tau_x = 2 \mu \frac{\partial{u}}{\partial{x}} + \mu \left[\frac{\partial{u}}{\partial{y}} + \frac{\partial{v}}{\partial{x}}\right] + \mu \left[\frac{\partial{u}}{\partial{z}} + \frac{\partial{w}}{\partial{x}}\right]
	\end{equation}
	
	As it is evident from Figure \ref{fig8}, flow for NPR=3 is attached, whereas, for NPR=6, it is separated. Lateral confinement keeps the flow features relatively two dimensional (separation line being straight), whereas free lateral expansion forms symmetrical jet distortions about the symmetry plane (separation line being curved). Separation lines are marked using the dotted yellow line in Figure \ref{fig8}-Ia and \ref{fig8}-Ib. The separation line is found to be closer to the throat when there is a free lateral expansion $ ([x/L]\approx0.75) $, whereas lateral confinement keeps the separation slightly away from the throat $ ([x/L]\approx0.8) $.
	
	Corresponding computations also reveal almost the same features. Streamlines drawn closer to the wall marks the distorted and separation zones. The separation zones could be readily identified in Figure \ref{fig8}-IIa and \ref{fig8}-IIb for NPR=6 by seeing the absence of streamlines around the trailing edge of the planar plug nozzle $ (0.7 \leq [x/L] \leq 1) $. An oblique shock forms at the point of separation as the nozzle flow is supersonic upstream. In the case with no lateral confinement, three-dimensional separation leads to the formation of curved shocks.
	
	\sk{For the considered NPR’s, the expansion fan from the nozzle lip and the surface comes in contact with the lateral free shear layer. Upon impinging the free boundary, the expansion fan is reflected as compression waves. The fluid from the jet periphery thus takes a curvilinear profile towards the center-line, leading to the observation of curved shocks. Such a behaviour is well seen through the streamlines drawn from the computational studies in Figure \ref{fig8}-IIa. A similar phenomenon is also documented in the compressible free jet studies on the planar rectangular nozzle jet\cite{Behrouzi2018}, rectangular nozzle jet with after deck\cite{Frate2011,Behrouzi2015,Stack2018}, supersonic wall-jet flows\cite{Kwak2016} and linear cluster plug nozzle jet\cite{Chutkey2018}.}

	The observation from pressure plots shown in Figure \ref{fig6} strongly suggests that there is a sudden rise in the wall pressure values at locations predicted by qualitative visualization from oil flow experiments and numerical results (shown in Figure \ref{fig8}). The rise in pressure values is mainly due to the presence of over-expansion shockwaves on the pug surface. These shockwaves interact with the growing boundary layer. The interaction results in the thickening of the boundary layer and eventually leads to flow separation and results in the formation of a recirculation bubble on its surface. The portion on the plug surface where the shock interacts with the boundary layer is hereafter called the interaction region.
	
	The oil flow visualizations reveal that the flow remains attached for NPR=3, irrespective of the lateral confinement. On the contrary, for NPR=6, a recirculation bubble can be seen due to separation. The absence of lateral confinement makes the separation further three-dimensional. The shock wave on the plug surface for NPR=3 might not be strong enough to cause the flow separation; instead, the boundary layer thickens due to the presence of an adverse pressure gradient. 
	
	The flow Mach number variation along the center-line and side-line of the plug surface was evaluated using isentropic relation until the shock location on the plug surface. An increment in the magnitude of Mach number from 1.38 to 1.83 is observed at the center-line with side-wall as NPR increases from 3 to 6. Due to this increase in Mach number, the shock strength is enhanced, and the interaction with the boundary layer is stronger, thereby resulting in a larger interaction region and a larger recirculation bubble for NPR=6. Additionally, flow acceleration near the side-wall due to the displacement effect of the side-wall boundary layer aids in making the flow more resistant to separation \cite{SUDANI_1993,Su_1989}.
	
	\subsection{Pressure distribution}
	
	\subsubsection{Center-line pressure measurements}
\begin{figure*}
		\centering
		\includegraphics[width=0.8\textwidth]{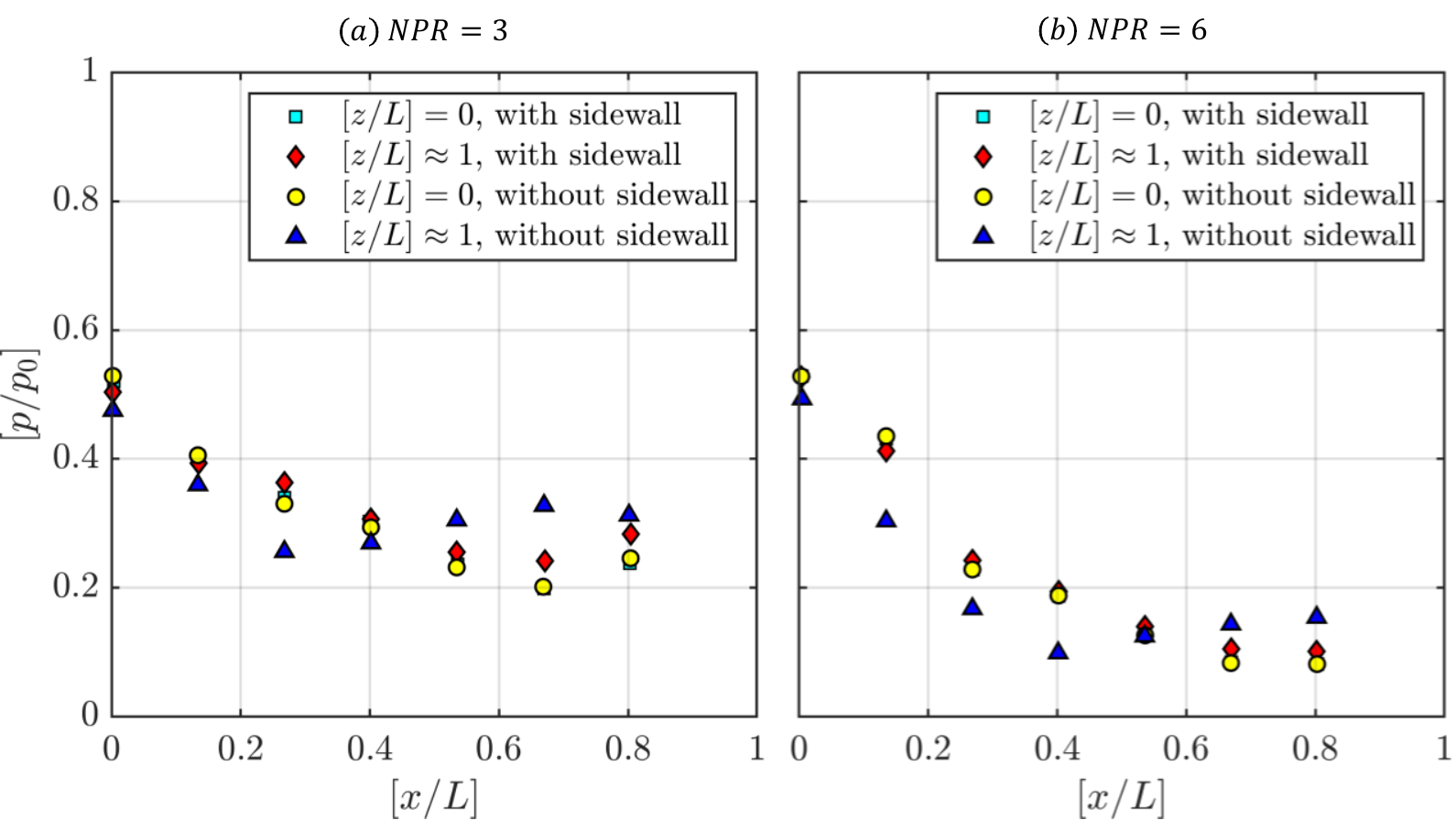}
		\caption{Normalized wall-static pressure distribution $(p/p_0)$ along plug surface $ (x/L) $ at two different NPR's: (a) NPR=3, and (b) NPR=6 with and without side-walls along the center-line $ ([z/L]=0) $ and close to the side-wall $([z/L]\approx 1)$.}\label{fig9}
	\end{figure*}

	Figure \ref{fig9} depicts the wall-static pressure measurement over the plug surface in which $ [x/L]=0 $ represents the position of the throat, and the x-axis is non-dimensionalized by the slant-height of the plug surface $ (L) $. Along the center-line of the plug surface, $ ([z/L]=0) $, the normalized wall-static pressure distribution $ (p/p_0) $ is invariant among the cases with and without side-wall for both the NPR's under investigation. 
	
	Values of $ p/p_0 $ decrease along the flow direction due to the expansion of the flow until the location $ [x/L] = 0.67 $, after which a rise in pressure was observed at $ [x/L] = 0.8 $, which is the last available port for pressure measurements. Somewhere between $ [x/L] = 0.67 $ and 0.8, there exist a region where a sudden pressure jump should be occurring. Based on the schlieren and oil flow visualization, we observe that this pressure rise is achieved through over-expansion oblique shockwave. This mechanism exists to equalize the expanding gas pressure $ (p) $ to the ambient pressure $ (p_a) $. 
	
	Unlike the NPR=3 case, for NPR=6, no significant rise in pressure values were observed along the plug surface; however, schlieren visualization (both experimental and numerical) showed a strong oblique shock wave. As a result of the higher strength of this over-expansion shock, more rise in static pressure was expected; nevertheless, the presence of interaction region (Figure \ref{fig8}) and the characteristic altitude adaptability of the pug nozzle played a vital role in decreasing the degree of increment in pressure, thereby less rise in pressure was perceived as can be seen from Figure \ref{fig6}. 
	
	\subsubsection{Sideline pressure measurements} 
	Along the side-line at $ [z/L]\approx 1 $ in the presence of the side-wall, the pressure distribution is different (Figure \ref{fig9}). It is slightly higher when compared to the measurements at $ [z/L]=0 $, owing to the developing boundary layer on the plug surface and the side-wall. However, the pressure distribution trend and the location of the oblique shock wave on the plug surface are similar to those along the center-line $ ([z/L]=0) $. Such observations also confirm that the separation line formed due to the oblique shock wave along the spanwise direction is straight, suggesting the flow field along the plug surface remains almost two dimensional in the absence of the side-walls at NPR=6. In the absence of a side-wall, there is a rapid expansion of jet in the lateral direction starting from the throat region. This is, in turn, confirmed by the pressure measurements (see Figure \ref{fig9}) along the side-line at $ [z/L]\approx 1 $, which showed a huge difference in pressure values compared to the case with the side-wall being present. The separation line thus cannot be two-dimensional for the case without the side-wall. The interaction is complex and three-dimensional. Moreover, in the absence of side-wall, along the sideline of the plug surface, the pressure jump is around $ [x/L] = 0.27 $ for NPR=3 and around $ [x/L] = 0.4 $ for NPR=6, in contrary to the pressure rise along the center-line $ ([z/L]=0) $. This intuitively suggests that the separation line due to the oblique shock wave is curved due to lateral expansion.
	
	\begin{figure*}
		\centering
		\includegraphics[width=0.9\textwidth]{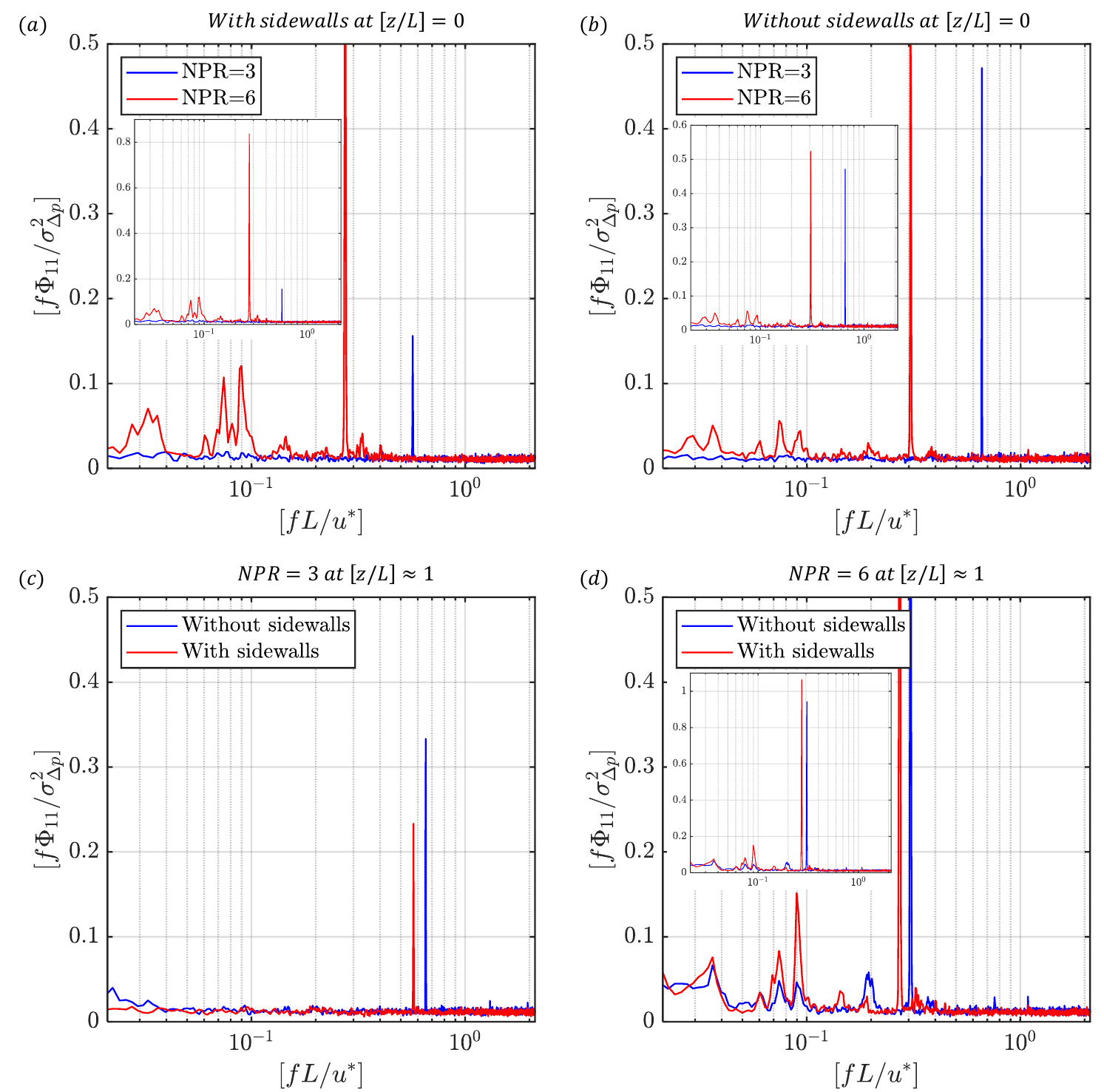}
		\caption{\sks{Non-dimensionalized pre-multiplied power spectra of the wall-static pressure fluctuations $ (\Delta p) $ from the unsteady PCB sensors placed on the planar plug nozzle surface at different NPR's (NPR=3 and NPR=6, a-b), along the  (a, c) center-line (at the location $ [x/L=0.8] $ and $ [z/L=0] $) and (b, d) close to the side-wall (at the location $ [x/L=0.8] $ and $ [z/L\approx 1] $) for both the cases: with and without side-walls (c-d). A common variance of $ \sigma_{\Delta p}^2= 1 \times 10^4$ Pa$^2$ is used for normalization to appreciate the comparison of power spectra. Similarly, frequency in $x$-axis is normalized using the throat velocity ($u^*$) and plug nozzle surface length ($L$), as they remain constant for all the cases under consideration.}}\label{fig10}
	\end{figure*}
	
	\subsection{Power spectra}
	The flow field on the plug surface is observed to be unsteady from both schlieren and oil flow visualization. \sk{In order to study the temporal characteristics of the dynamic pressure signal in the interaction region, a spectral analysis is carried out using the \textit{pwelch} method (Hamming window with 50\% overlap between segments) using MATLAB\textsuperscript{\tiny\textregistered} to deduce the frequency spectrum and the associated Power Spectral Density (PSD).} Since the objective of the present investigation is to study the effect of lateral confinement on plug surface flow field, the spectral analysis is carried out only for two locations inside the interaction region: along the center-line $ ([z/L]=0) $ and closer to the side-wall $ ([z/L]\approx 1) $ at $ [x/L]=0.8 $. 
	The pre-multiplied \textit{PSD} is normalized using a common variance for the comparison of different cases. The obtained spectra for NPR=3 and 6, at the center-line $ [z/L]=0 $ with the lateral confinement, is plotted in Figure \ref{fig10}a. 
	
	The spectra show a notable difference, especially in the low-frequency range. The spectra do not show any significant spectral feature except for a dominant peak around 8 kHz for NPR=3, where the flow remains attached to the plug surface. On the other hand, NPR=6, where the flow is separated due to shockwave-boundary layer interaction, has shown several spectral features, $\sim 10^3$ Hz ($fL/u^* \sim 0.0707$). Two significant broadband spectra were observed, the first one between 0.35 – 0.48 kHz (0.025 - 0.034) with a peak value at 0.45 kHz and the second one between 1 – 1.6 kHz (0.07 - 0.113) with dominant peaks at 1.1 kHz (0.078) and 1.5 kHz (0.106). While the low-frequency large-amplitude fluctuations (0.45 kHz, $fL/u^*=0.032$) can be attributed to the oscillation of separation shock foot \cite{Verma_2014}, the high-frequency high-amplitude fluctuation (1 – 1.6 kHz) can be due to the motions induced by shock-shear layer interactions \cite{Olson_2013,Chen_2018}. \sk{However, the amplitude of the aforementioned spectra is minimal here in comparison with the dominant tone observed around 4 kHz (0.3). The reason is primarily because of the different nozzle configuration (conventional C-D nozzle \cite{Olson_2013}). The influence of screeching or the production of distinct high-frequency sound pulse/spherical wave-front, occurring outside the nozzle due to jet-flapping, is attenuated considerably due to the bounding walls of the nozzle walls unlike the plug nozzle. However, in a plug nozzle, most of the jet boundary is exposed to the atmosphere which can be subjected to the perturbation of screeching.} 
	
	Other spectral features observed for NPR=6 are bands centered around 2 kHz (0.141) and 4.5 kHz (0.318); however, their amplitudes are very less when compared to others. In the absence of lateral confinement, the spectral features remained the same for both the NPR's except that their amplitude is smaller, as shown in Figure \ref{fig10}b. Irrespective of the lateral confinement, a prominent peak of equivalent power is observed at 0.85 kHz (0.0601) for NPR=6. \sk{For NPR=3 the strength of the shock is weaker when compared with NPR=6, hence, its intensity or energy level is very minimal when compared with its other dominant spectral feature and those of NPR=6.} Earlier works on the plug nozzle also reported a similar peak at the same frequency and have associated it with the over-expansion shock \cite{Verma_2011,Chutkey_2017}. \sk{However, the researchers utilized a linear cluster of plug nozzles which is moderately different in comparison to the one used in our present experiments. \sks{Linear clusters offer shielding in a similar way to that of a side-wall where the associated flow physics partially differ from the present one.} In addition, the researchers used an extended cowl where part of the jet-expansion had already happened internally. In the present case, flow expansion begins from the throat itself (freely on one side and bounded by the plug nozzle surface on the other side). Besides, the high frequency oscillations\cite{Verma_2011,Chutkey_2017} are only prominent at NPR 8.8 (Fig 5b) and not at the present operating conditions under discussion. Apart from the different NPR conditions, there was a strong base flow which provides further shielding and additional flow features, especially the reduced percentage of exposed separation zone prone to screech effects at lower NPR.}
	
	\sk{Similar to NPR=3 having a dominant peak around 8 kHz ($fL/u^*=0.5654$), NPR=6 also exhibits a dominant peak, however, it is observed at 3.85 kHz ($fL/u^*=0.272$).} Similar observations were also reported in the earlier research works on plug nozzles \cite{Verma_2011,Chutkey_2017}, where they observed dominant frequency at 3.36 kHz (0.2375) for NPR$\approx$11 and at 3.8 kHz (0.2686) at NPR$\approx$9. These high frequencies are attributed to the overall unsteadiness on the plug surface, and it is reported that the entire flow was fluctuating at this single frequency. However, from our present studies, we speculate that these frequencies are due to acoustic screech. The acoustic radiation generated downstream of the nozzle exit propagates upstream and interacts with the thin shear layer at the nozzle lip. \sks{While doing so, the upstream travelling disturbances intensify the instabilities in the shear layer. The amplified instabilities from the shear layer are further convected downstream and contribute to the production of acoustic radiation. The generated acoustic radiation then again propagates upstream, thereby forming a feedback loop and establishes the basis for discrete frequency generation by the jet flow \cite{Tam1995}.} Unlike conventional C-D nozzle, this interaction of acoustic radiation with the shear layer occurs at the throat exit, due to the absence of plug on the upper throat-lip, thereby influencing the fluctuations on the plug surface in the bottom throat-lip. This feature of acoustic screech influencing the fluctuation on the plug surface seems to be a characteristic of plug nozzle.
	
	\sk{The dominant frequency in Figure \ref{fig10}b is found to decrease from 9.3 kHz (0.6573) to 4.35 kHz (0.3074) with an increase in NPR from 3 to 6.} The behavior of lower-shifting of the peak frequency with an increase in NPR is similar to that of the events observed in the free jets \cite{Chen_2018}. Notably, the screeching frequency is attributed to the \textit{Mode-B} events observed at higher frequency kHz) (between 4 to 8 kHz) when the NPR is between 3-7. 
	
	\begin{figure*}
		\centering
		\includegraphics[width=0.8\textwidth]{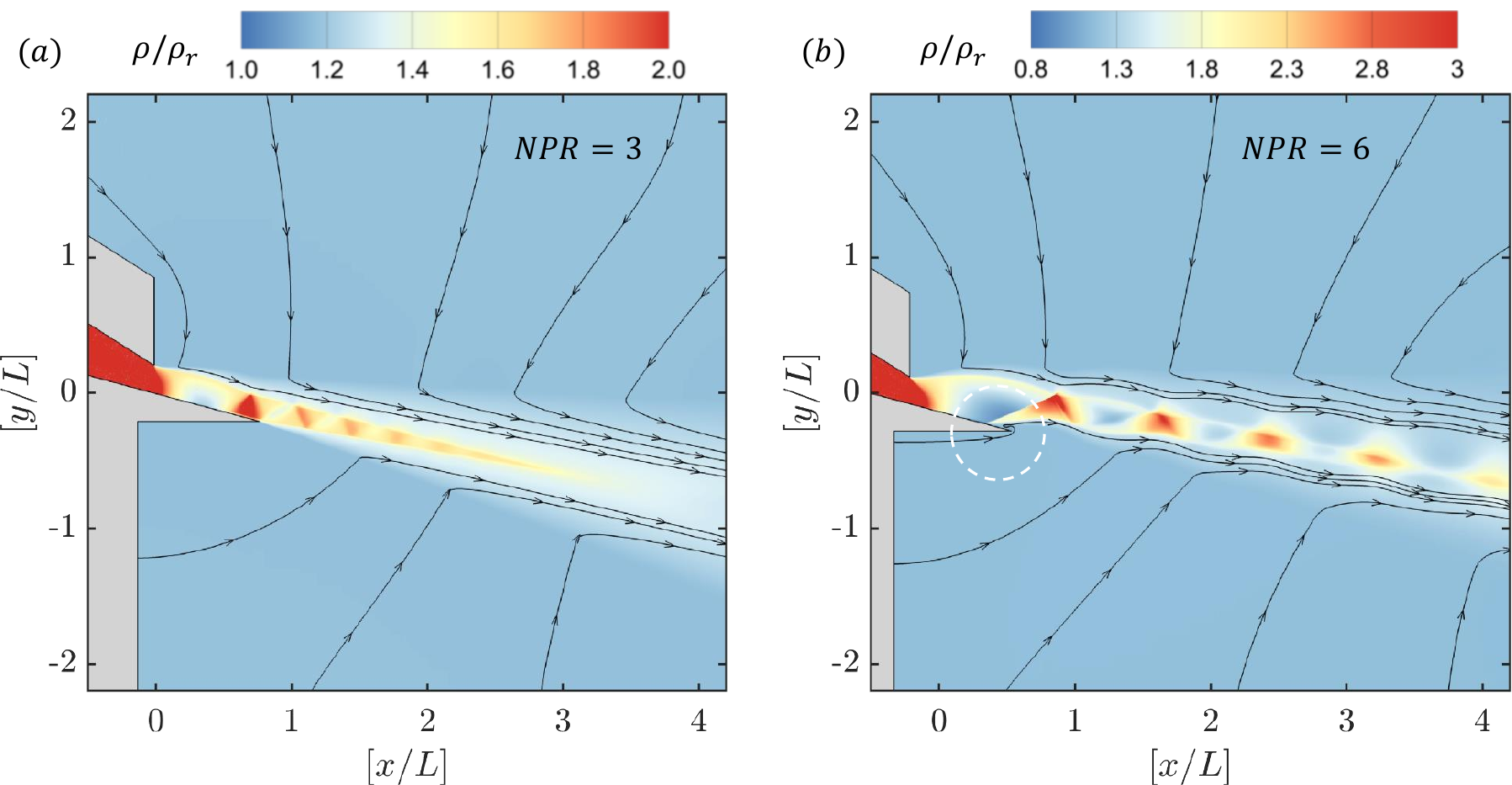}
		\caption{Two-dimension contour plot of the non-dimensionalized density ($\rho/\rho_r$, where $\rho_r$=1 kg/m$^3$) in the $xy$-plane at $ [z/L=0] $ at different NPR's: (a) NPR=3, and (b) NPR=6 for both the two-dimensional case which resembles the case with side-walls. The dotted white-circle points the passage of an almost parallel streamline through the separated flow region on the planar plug nozzle surface.}\label{fig11}
	\end{figure*}
	
	The spectra from the unsteady pressure signal measured along the side-line $ ([z/L]\approx 1) $, in the presence and absence of side-wall, for different NPR's are shown in Figure \ref{fig10}c-d. For both the NPR's, no noticeable changes in the spectral lines are observed due to the lateral confinement in comparison with the spectra seen at the center-line $ ([z/L]=0) $, except a slight power drop. However, the power of the dominant frequency for NPR=6 is increased significantly in the absence of the side-wall. For NPR=6, the spectral features are similar to those observed at the center-line; however, in the absence of side-wall, their amplitudes in the frequency range of 1 – 1.5 kHz (0.071 - 0.106) are noticeably lower owing to the sideways spillage of the separated flow. On the other hand, the spectra had the same amplitude at 0.4 kHz (0.0283) and 0.85 kHz (0.0601) both in the presence and absence of the side-wall. 
	
	Another noticeable event includes the influence of the screech on the interaction region. Even though the flow is attached in NPR=3, when the lateral confinement is absent, the flow is forced by the screech frequency. The influence is seen in Figure \ref{fig10}c, where the power is almost twice the case with lateral walls. In addition, a small positive shift in frequency is seen for both the operating conditions. The downstream acoustic forcing of the upstream screech through the subsonic portion of the boundary layer might create such narrow spectra at the point of measurement. 
	
	From the computational results in Figure \ref{fig11}, the communication or forcing from the environment back to the interaction region is seen between the two different NPR conditions. At NPR=3, the flow is observed to be attached, and the separated shear layer is comparatively thinner than the NPR=6 case. At NPR=6, due to flow separation on the ramp wall surface, the streamlines from the ambient pass through the separation zone (the region is marked as a dotted white circle). \sks{In the absence of lateral confinement for different NPR's, tonal disturbances from the screech \textit{(Mode-B)} severely influence the unsteady dynamics around the flow separation zone, which explain the varying power levels observed on the dominant frequency between NPR=6 and NPR=3 at $[z/L\approx 1]$ in Figure \ref{fig10}c-d.}
	
	\begin{figure*}
		\centering
		\includegraphics[width=0.9\textwidth]{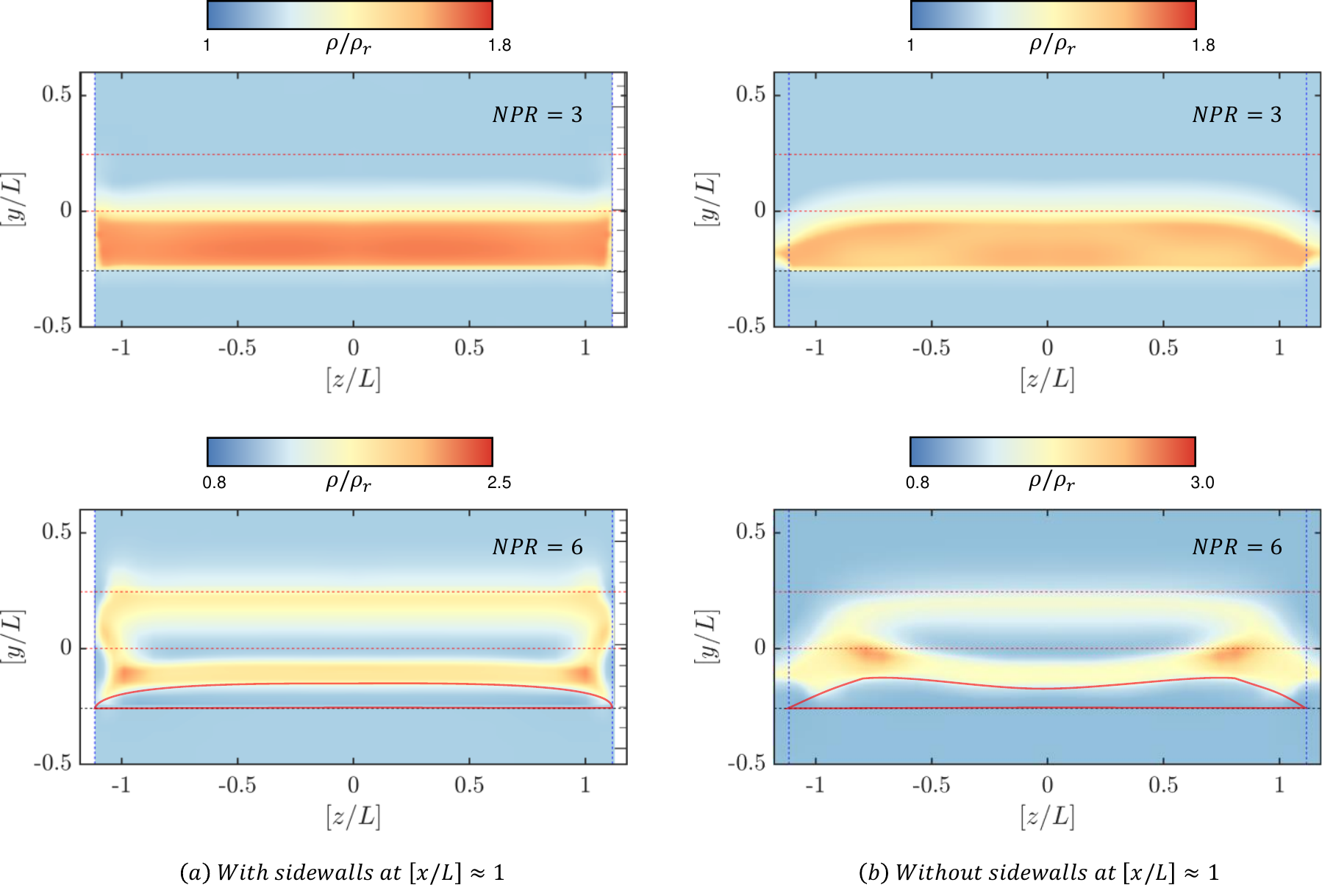}
		\caption{Two-dimension contour plot of the non-dimensionalized density ($\rho/\rho_r$, where $\rho_r$=1 kg/m$^3$) in the $ yz $-plane at $ [x/L\approx 1] $ at different NPR's (NPR=3 and NPR=6) for both the cases (a) with side-walls and (b) without side-walls. The horizontal dotted red-lines mark the locus of the upper $ [y/L]\approx 0.25 $ and lower edge $ [y/L]=0 $ of the throat. The horizontal dotted black-line mark the edge of the plug surface $[y/L] \approx -0.25$ lip at $[x/L \approx 1]$. The vertical dotted blue-lines mark the limiting span $ (w) $ of the planar plug nozzle $(-1.1\leq [z/L] \leq 1.1)$. The solid red line marked at NPR=6 case represents the region of separated flow between the plug surface and the lower jet boundary.}\label{fig12}
	\end{figure*}
	
	\subsection{Three-dimensional effects}
	As evident from the previous measurements, three-dimensional events are prominent in the absence of lateral confinement. Computational results from the transverse planes are analyzed to assess the extent of the three-dimensional flow features. Slices are taken at $ [x/L]\approx 1 $ in all the cases, and the normalized density contours are visualized. From Figure \ref{fig12}, the flow seems two-dimensional only for NPR=3 under lateral confinement. For higher NPR, three-dimensional features start to propagate from the walls to the core. The separation is predominant in the core flow, and the jet remains attached closer to the wall.
	On the contrary, the jet structure is three dimensional for both the NPR's when there is a free lateral expansion. However, the severity is more for NPR=6 cases. A visible distorted mouth-like structure could be seen with a strong lifted-up vortex closer to the extrema of the nozzle. The effective jet diameter leaving the plug nozzle under such strong three dimensional interactions would be much less as seen in Figure \ref{fig12}b. These changes alter the near and far-field jet characteristics, in addition to the variations observed on the side load characteristics. 
	
	\section{Conclusions} \label{conclusions}
	A planar plug nozzle with and without side-walls are analyzed both numerically and experimentally at two different over-expanded nozzle pressure ratio (NPR=3 and 6). Schlieren and oil flow visualization are used as qualitative tools, whereas the steady and unsteady pressure measurements are utilized to extract quantitative information. Computations are identified to replicate results closer to experiments. Experimentally inaccessible flow planes are visualized computationally to extract flow physics. Following are the major conclusions from our study:
	\begin{itemize}
		\item Lateral confinement keeps the separation at higher NPR fairly two-dimensional. Upon free lateral expansion, the separation zones are observed to be larger and three-dimensional.
		\item At higher NPR's, unsteady flow separation is severe and observed to be in the low-frequency range. 
		\item \sk{The frequency contents prominently shift to higher values due to the presence of confinement irrespective of NPR's. However, during free lateral expansion when there is no confinement, the power of the spectrum drops drastically.}
		\item Lateral confinement offers shielding from the downstream screech, which reduces the power to almost half of the value.            
	\end{itemize}
	
	\section*{Author's Contributions}
	All authors have contributed equally to this work. 
	
	\section*{Data Availability}
	The data that support the findings of this study are available from the corresponding author upon reasonable request. 
	
	\section*{Acknowledgments}
	The authors would like to thank Mr. Rohit Panthi, Mr. Shishupal Singh Sachan, and Mr. Suresh Mishra for their help during the experimental work. The financial support from the Department of Aerospace Engineering, Indian Institute of Technology, Kanpur, for carrying out the research work is gratefully acknowledged. \sk{The authors also acknowledge the support of Indian Institute of Technology Kanpur (IITK) (www.iitk.ac.in/cc) for providing the resources to carry out the simulation, and data analysis as well. The authors gratefully acknowledge the useful comments of the reviewers in shaping the manuscript.} 
	
	\bibliography{ibrahimPOF}
	
\end{document}